\documentclass[letter]{aa} 
\usepackage{graphicx}
\usepackage{natbib,twoopt}
\usepackage{flushend}
\usepackage[colorlinks,linkcolor=blue,urlcolor=blue,citecolor=black]{hyperref} 
\bibpunct{(}{)}{;}{a}{}{,}             
\makeatletter
  \newcommandtwoopt{\citeads}[3][][]{\href{http://ui.adsabs.harvard.edu/abs/#3}%
    {\def\hyper@linkstart##1##2{}%
     \let\hyper@linkend\@empty\citealp[#1][#2]{#3}}}
  \newcommandtwoopt{\citepads}[3][][]{\href{http://ui.adsabs.harvard.edu/abs/#3}%
    {\def\hyper@linkstart##1##2{}%
     \let\hyper@linkend\@empty\citep[#1][#2]{#3}}}
  \newcommandtwoopt{\citetads}[3][][]{\href{http://ui.adsabs.harvard.edu/abs/#3}%
    {\def\hyper@linkstart##1##2{}%
     \let\hyper@linkend\@empty\citet[#1][#2]{#3}}}
  \newcommandtwoopt{\citeyearads}[3][][]%
    {\href{http://ui.adsabs.harvard.edu/abs/#3}
    {\def\hyper@linkstart##1##2{}%
     \let\hyper@linkend\@empty\citeyear[#1][#2]{#3}}}
  \renewcommand*\aa@pageof{, page \thepage{} of \pageref*{LastPage}} 
\makeatother
\usepackage{txfonts}
\begin{document}

\title{A stripped helium star in the potential black hole binary LB-1
}

\author{A.~Irrgang\inst{\ref{remeis}}
                \and
                S.~Geier\inst{\ref{potsdam}}
                \and
                S.~Kreuzer\inst{\ref{remeis}}
                \and
                I.~Pelisoli\inst{\ref{potsdam}}
                \and
                U.~Heber\inst{\ref{remeis}}
       }
       
\institute{
Dr.~Karl~Remeis-Observatory \& ECAP, Astronomical Institute, Friedrich-Alexander University Erlangen-Nuremberg (FAU), Sternwartstr.~7, 96049 Bamberg, Germany\\
\email{andreas.irrgang@fau.de}\label{remeis}
\and
Institut f\"ur Physik und Astronomie, Universit\"at Potsdam, Karl-Liebknecht-Str.\ 24/25, 14476 Potsdam, Germany\label{potsdam}
}
\date{Received 18 December 2019 / Accepted 1 January 2020}

\abstract{The recently claimed discovery of a massive ($M_\textnormal{BH}=68^{+11}_{-13}\,M_\odot$) black hole in the Galactic solar neighborhood has led to controversial discussions because it severely challenges our current view of stellar evolution.}
{A crucial aspect for the determination of the mass of the unseen black hole is the precise nature of its visible companion, the B-type star LS\,V$+$22\,25. Because stars of different mass can exhibit B-type spectra during the course of their evolution, it is essential to obtain a comprehensive picture of the star to unravel its nature and, thus, its mass.}
{To this end, we study the spectral energy distribution of LS\,V$+$22\,25 and perform a quantitative spectroscopic analysis that includes the determination of chemical abundances for He, C, N, O, Ne, Mg, Al, Si, S, Ar, and Fe.}
{Our analysis clearly shows that LS\,V$+$22\,25 is not an ordinary main sequence B-type star. The derived abundance pattern exhibits heavy imprints of the CNO bi-cycle of hydrogen burning, that is, He and N are strongly enriched at the expense of C and O. Moreover, the elements Mg, Al, Si, S, Ar, and Fe are systematically underabundant when compared to normal main-sequence B-type stars. We suggest that LS\,V$+$22\,25 is a stripped helium star and discuss two possible formation scenarios. Combining our photometric and spectroscopic results with the {\it Gaia} parallax, we infer a stellar mass of $1.1\pm0.5\,M_\odot$. Based on the binary system's mass function, this yields a minimum mass of $2$--$3\,M_\odot$ for the compact companion, which implies that it may not necessarily be a black hole but a massive neutron- or main sequence star.}
{The star LS\,V$+$22\,25 has become famous for possibly having a very massive black hole companion. However, a closer look reveals that the star itself is a very intriguing object. Further investigations are necessary for complete characterization of this object.}


\maketitle
\section{Introduction \label{sec:intro}}
Stellar black holes are hard to find if they are not in interacting binaries, that is, in X-ray binaries. Recently, \citetads{2018MNRAS.475L..15G} found a single-lined binary in the globular cluster NGC\,3201 consisting of a turn-off star orbited by a detached compact object of minimum mass $4.36\pm0.41\,M_\odot$ in a highly eccentric orbit with a period of $166.8$\,d. Because this minimum mass exceeds the Oppenheimer-Volkoff limit, the discovery represents the first direct mass estimate of a detached black hole in a globular cluster.

Most recently, \citetads{2019Natur.575..618L} reported the discovery of the long-period (single-lined) spectroscopic binary system LB-1 (LS\,V$+$22\,25). The visible component of this system is a bright B-type star with a prominent and broad H$\alpha$ emission line very similar to a Be star. A fit of TLUSTY B-star model spectra lead to an effective temperature $T_{\textnormal{eff}} = 18\,104\pm825$\,K, a surface gravity of $\log(g)=3.43\pm0.15$, and a metallicity consistent with the solar value. Comparing these atmospheric parameters with stellar evolution models of B-type main sequence (MS) stars, the authors concluded that it is a young, massive star with $M=8.2^{+0.9}_{-1.2}\,M_\odot$ and $R=9\pm2\,R_\odot$ located in the direction of the Galactic anti-center at a distance of $d=4.23\pm0.24$\,kpc.

The absorption lines of the star show radial-velocity variations with a period $P=78.9\pm0.3$\,d. Fitting a binary orbit to the radial-velocity curve yielded a semi-amplitude $K_1=52.8\pm0.7$\,km\,s${}^{-1}$ and a very small eccentricity $e=0.03\pm0.01$. Adopting the above-mentioned mass for the visible B-star, the mass of the companion would have to be higher than $6.3^{+0.4}_{-1.0}\,M_\odot$. Since no spectral features of another star are detectable in the spectra, \citetads{2019Natur.575..618L} concluded that the companion cannot be a star, but instead has to be a black hole. 

Based on the complex shape and substantial width ($\sim240$\,km\,s${}^{-1}$) of the H$\alpha$ emission line, the authors concluded that it can only be a Keplerian disk. Since its radial velocity does not follow the motion of the star, but shows a very small variation ($K_\alpha=6.4\pm0.8$\,km\,s${}^{-1}$) in anti-phase, the authors propose that the disk is connected to the black hole companion. If so, the orbital inclination of the binary would be very small ($i_\textnormal{o}=15$--$18^{\circ}$) and the black hole extremely massive ($M_\textnormal{BH}=68^{+11}_{-13}\,M_\odot$). This result strongly challenges stellar evolution models, which do not predict black holes of such high masses to be formed in metal-rich environments (\citeads{2019Natur.575..618L}; \citeads{2019arXiv191203599E}; \citeads{2019arXiv191200994G}; \citeads{2019arXiv191112357B}). The conclusion of \citetads{2019Natur.575..618L} rests on two main pillars: first, the B-star is a massive MS star; and second, the emission line is connected to the compact companion. However the latter was severely questioned by \citetads{2019arXiv191204092A} and \citetads{2019arXiv191204185E} who argue that the radial velocity variation of the emission line is actually caused by the variability of the underlying absorption line of the visible star and is therefore not related to the compact companion. Because the H$\alpha$ emission appears to be stationary, it might result from a circumbinary disk. Consequently, the measured mass function
\begin{equation}
f(M) \coloneqq \frac{M_2 \sin^3(i_\textnormal{o})}{(1+M_1/M_2)^2} = (1-e^2)^{3/2}\frac{K_1^3 P}{2 \pi G} =  1.20\pm0.05\,M_\odot \,,
\label{eq:mass_function}
\end{equation}
seems to remain as the only constraint on the mass $M_2$ of the compact companion, which is then a function of two unknowns, namely the mass of the visible star $M_1$ and the orbital inclination $i_\textnormal{o}$. The spectral analysis by \citetads{2019arXiv191204092A} revised the atmospheric parameters to $T_{\textnormal{eff}} = 13\,500\pm700$\,K, $\log(g)=3.3\pm0.3$, and a projected rotational velocity of $\varv\sin(i)=7.5\pm4.0$\,km\,s${}^{-1}$. Those parameter values might still be consistent with a massive B-type subgiant. However, the very low projected rotational velocity --~although it can be explained by seeing the object relatively pole-on~-- is quite unusual for such an object and could also be indicative of a different nature. Indeed, some evolved low-mass stars are known that can mimic normal B-type stars spectroscopically (see, e.g., \citeads{1997ApJS..111..419H}; \citeads{2001A&A...379..235R}); these can be distinguished from massive B stars by their slow rotation and chemical anomalies. Therefore, in order to unravel the nature of the visible star and hence its mass, we embarked on a quantitative spectroscopic analysis of LS\,V$+$22\,25. This analysis shows that this object is very likely a rare stripped helium star in an evolutionary stage, at which it simply mimics a massive MS star. This has important consequences for the mass and potentially also for the nature of its compact companion.
\section{Analysis}
\subsection{Quantitative spectroscopic analysis}
\begin{figure}
\centering
\includegraphics[width=0.49\textwidth]{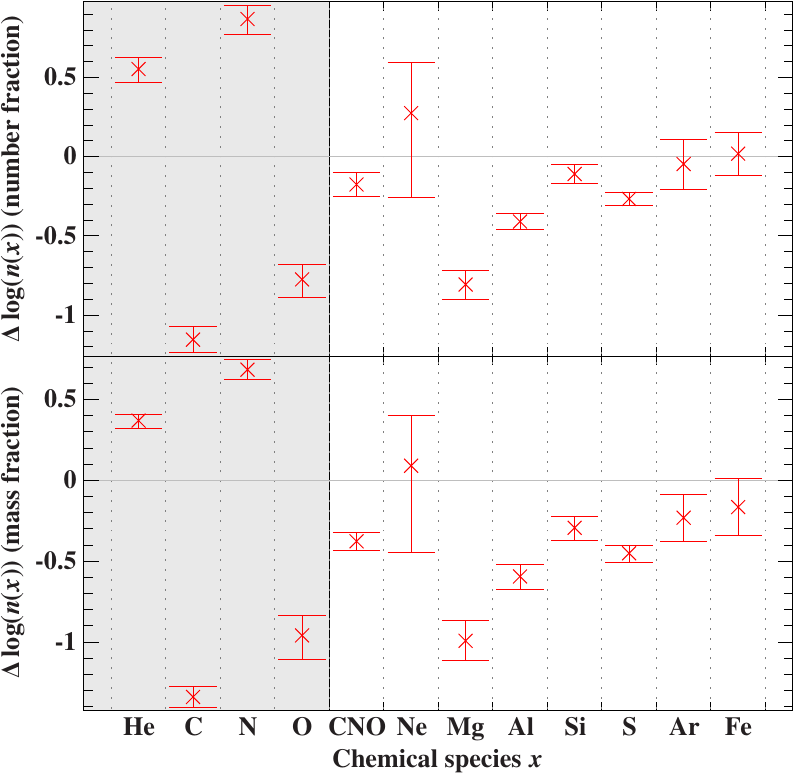}
\caption{Differential abundance pattern of LS\,V$+$22\,25 with respect to the cosmic abundance standard by \citetads{2012A&A...539A.143N} in terms of fractional particle numbers (\textit{upper panel}) and mass fractions (\textit{lower panel}). Solar abundances by \citetads{2009ARA&A..47..481A} are used for Al, S, and Ar because cosmic abundances are not available for these elements. The error bars are the square roots of the quadratic sums of all given individual $1\sigma$ uncertainties. The gray-shaded area marks elements whose abundances show signatures of hydrogen burning via the CNO bi-cycle. To get rid of this effect, the invariant sum of the bi-cycle-catalysts C, N, and O (``CNO'') is also plotted. The abundance pattern of metals heavier than oxygen is nonstandard as well.}
\label{fig:abundance_pattern}
\end{figure}
Our spectral investigation is based on the publicly available high-resolution Keck/HIRES spectra described by \citetads{2019Natur.575..618L}. The analysis strategy and the applied models are explained in detail in \citetads{2014A&A...565A..63I}. Briefly, model computations follow a so-called hybrid approach in which the structure of the atmosphere is computed in local thermodynamic equilibrium (LTE) with {\sc Atlas12} \citepads{1996ASPC..108..160K} while departures from LTE are accounted for by applying updated versions of {\sc Detail} and {\sc Surface} (\citeads{1981PhDT.......113G}; \citealt{detailsurface2}). The {\sc Detail} code computes population numbers in non-LTE by numerically solving the coupled radiative transfer and statistical equilibrium equations. The {\sc Surface} code uses the resulting departure coefficients and more detailed line-broadening data to compute the final synthetic spectrum. All three codes have been recently updated \citepads{2018A&A...615L...5I} by allowing for non-LTE effects on the atmospheric structure as well as by the implementation of the occupation probability formalism \citepads{1994A&A...282..151H} for hydrogen and new Stark broadening tables for hydrogen \citepads{2009ApJ...696.1755T} and neutral helium \citepads{1997ApJS..108..559B}. Moreover, the concept of opacity sampling is consistently used throughout all computational steps, which is important when atmospheres with nonstandard chemical compositions are analyzed. The best-fitting model parameters are found by a simultaneous fit of all available spectra over their entire useful spectral range; see Fig.~\ref{fig:spectra} for a comparison.

Our atmospheric parameters ($T_{\textnormal{eff}} = 12\,720\pm260$\,K, $\log(g)=3.00\pm0.08$, see Table~\ref{table:atmospheric_parameters}) are, within error bars, consistent with those of \citetads{2019arXiv191204092A} but quite different from those of \citetads{2019Natur.575..618L}. Furthermore, the low values for the projected rotational velocity and for the microturbulence, which were adopted or measured by \citetads{2019Natur.575..618L} and \citetads{2019arXiv191204092A}, are more or less confirmed here ($\varv\sin(i)=8.7\pm0.2$\,km\,s${}^{-1}$, $\xi\leq0.1$\,km\,s${}^{-1}$). While those values are already quite uncommon for a MS B-type star, the most striking result is revealed by the abundance analysis; see Fig.~\ref{fig:abundance_pattern}. The star shows an overwhelming signature of material that is processed by hydrogen burning via the CNO bi-cycle, namely extremely strong enrichment in helium and nitrogen at the expense of carbon and oxygen. Additionally, the heavier chemical elements also exhibit pronounced peculiarities. Magnesium and aluminum are considerably depleted while silicon, sulfur, argon, and iron are closer to ``standard'' values, but still systematically depleted in terms of mass fractions\footnote{We focus here on mass instead of number fractions because the former are invariant against stellar evolution, i.e., they do not change when hydrogen is converted to helium. Consequently, they are the natural measure for chemical peculiarities when stellar evolution is at play.}. Atomic diffusion processes could possibly explain such an abundance pattern, in particular because the observed helium lines show an unusual shape: the observed line wings are too strong while the line cores are too weak (see Fig.~\ref{fig:spectra_helium}). This could be the signature of vertical stratification due to diffusion in the atmosphere, as observed in the spectra of some blue horizontal branch and hot subdwarf B stars (sdBs, see \citeads{2018A&A...618A..86S} and references therein). However, diffusion is unlikely to act in a star with such low gravity ($\log(g)=3.00$). Therefore, it is more plausible to attribute the mismatch of the helium lines to disk emission filling the line cores, which is the same effect that is observed for the hydrogen Balmer lines \citepads{2019Natur.575..618L}. However, much higher temperatures would be required to see this effect also in helium. An improved understanding of the disk structure is needed to solve this open question. For the time being, we adopt rather conservative systematic uncertainties on temperature and gravity to compensate for this source of uncertainty.
\subsection{Analysis of the spectral energy distribution}
To cross-check our spectroscopic results and to obtain a more comprehensive picture of LS\,V$+$22\,25, we also fitted its spectral energy distribution using a two-component model consisting of a stellar spectrum plus a blackbody component. The latter was included to empirically account for the observed infrared excess, which might be associated with the disk structure. While the derived blackbody temperature ($T_\textnormal{bb}=1140^{+120}_{-100}$\,K) indicates that this thermal source is much cooler than the stellar component, we do not want to overinterpret this result given that the disk is certainly much more complex than a simple blackbody. Nevertheless, by including this component, we are able to reproduce the entire spectral energy distribution from the near-ultraviolet to the far-infrared (see Fig.~\ref{fig:photometry_sed}) yielding results (see Table~\ref{table:photometry_results}) that are consistent with spectroscopy. Another notable outcome of this exercise is an estimate for the star's angular diameter and for the interstellar reddening in its line of sight. Using the extinction law by \citetads{1999PASP..111...63F}, we obtain an extinction coefficient $R_V = 3.64^{+0.22}_{-0.20}$ and a color excess $E(B-V)=0.397\pm0.024$\,mag. Those values become important below, when the reliability of the {\it Gaia} parallax is discussed.
\section{The nature of LS V+22 25 and its unseen companion}
\subsection{B-type stars: not only a MS phenomenon}  
B-type spectra can be observed for stars of very different evolutionary stages. The most common ones are young, massive MS stars. However, the much less massive ($\sim0.47\,M_\odot$) and smaller ($\sim0.1$--$0.3\,R_\odot$) sdBs show relatively similar spectra. Most of them are likely core helium-burning stars located on or somewhat evolved away from the horizontal branch (\citeads{1993ApJ...419..596D}; \citeads{2002MNRAS.336..449H}). However, in addition, even less massive ($\sim0.2$--$0.47\,M_\odot$) helium stars, which are direct progenitors of helium white dwarfs (He-WDs) can show B-type spectra for a period of time ($\sim10^{6}$--$10^{7}$\,yr; \citeads{1998A&A...339..123D}; \citeads{2013A&A...557A..19A}; \citeads{2014A&A...571L...3I}) during their evolution (e.g.\ \citeads{2003A&A...411L.477H}; \citeads{2016A&A...585A.115L}). Those objects are closely related to the cooler and less massive extremely low-mass white dwarfs, which have been discovered in substantial numbers in recent years (\citeads{2016ApJ...818..155B}; \citeads{2019MNRAS.488.2892P}).

Most low-mass B-type stars are much less luminous than their MS counterparts and their higher surface gravities lead to a stronger broadening of the spectral lines. Moreover, the majority of them also show chemical peculiarities in their atmospheres due to earlier mixing episodes, diffusion, isotopic anomalies, or stratification (e.g.,\ \citeads{2006A&A...452..579O}; \citeads{2013A&A...549A.110G}; \citeads{2018A&A...618A..86S}). Finally, low-mass B-type stars are usually intrinsically very slow rotators \citepads{2012A&A...543A.149G}. Consequently, such objects can in general be easily distinguished from MS stars.
\subsection{Helium stars}  
One common feature of all sdB types is that they are helium stars, which is difficult to explain using evolutionary scenarios involving only single stars. The helium core of a red giant has to be stripped either well before or at the moment at which helium-burning starts in the core. The stripping is done either by stable Roche lobe overflow (RLOF) to a companion star or via unstable common-envelope ejection (\citeads{2002MNRAS.336..449H}; \citeads{2003MNRAS.341..669H}). Depending on the mass-transfer mechanism, very short-period post-common-envelope systems consisting of sdBs with small companions ($P\sim0.05$--$30$\,d, e.g.,\ \citeads{2015A&A...576A..44K}; \citeads{2019A&A...630A..80S}) or wide post-RLOF systems ($P\sim300$--$1200$\,d, \citeads{2013A&A...559A..54V}) are formed.

Only a few candidates for core helium-burning sdB stars have been proposed to have neutron star or black hole companions \citepads{2010A&A...519A..25G}, but an extended survey for such objects in close orbits could not confirm their existence (\citeads{2011A&A...530A..28G}; \citeads{2015A&A...577A..26G}). However, at least one pre-He-WD with a neutron star companion is known \citepads{2013ApJ...765..158K} and the more evolved He-WDs are regularly detected as visible companions of neutron stars. 

Stripped helium stars with masses higher than $\sim0.47\,M_\odot$ are also predicted by binary evolution models \citepads{2018A&A...615A..78G}. However, these are much rarer and only a few of them have been observed so far, such as for example HD\,49798 \citepads{2009Sci...325.1222M}. This enigmatic X-ray binary subdwarf O star (sdO) may be considered as the prototype for a stripped helium star \citepads{2017ApJ...847...78B}. Although HD\,49798 has been assigned the spectral type sdO, it is unique among the hot subdwarfs because it is considerably more massive and luminous. It is a single-lined spectroscopic binary with an orbital period of $1.5$\,d, which was discovered as a soft X-ray source by the ROSAT satellite. The presence of X-rays indicates that a compact object accretes from the stellar wind. X-ray eclipses allowed dynamical measurement of the masses of the two binary components, revealing that HD\,49798 is much more massive ($1.5\,M_\odot$) than normal hot sudwarfs stars ($\sim0.47\,M_\odot$) and that its companion is either a white dwarf or a neutron star of $1.28\pm0.05\,M_\odot$. HD\,49798 has been revisited recently, using both spectroscopy and evolutionary modeling. \citetads{2019A&A...631A..75K} carried out a quantitative spectral analysis of optical and ultraviolet spectra to derive the abundance pattern. Similar to LS\,V$+22$\,25, the atmospheric helium is strongly enriched and the CNO pattern is obvious. The abundances of heavier elements are also nonstandard: mostly overabundant with respect to the Sun. \citetads{2017ApJ...847...78B} computed the evolution of such a binary and were able to explain the observed abundances in helium, carbon, nitrogen, and oxygen as a result of stripping of the envelope of a $7\,M_\odot$ star by its compact companion.
\subsection{LS V+22 25: An intermediate-mass stripped helium star?}  
To constrain the mass, radius, and luminosity of LS\,V$+$22\,25, we make use of its angular diameter, surface gravity, and parallax from {\it Gaia} ($\varpi=0.4403\pm0.0856$\,mas), which places the star much closer than expected for a massive B-star. \citetads{2019Natur.575..618L} argued that the {\it Gaia} measurement could be erroneous because of the orbital motion of the star. However, the {\it Gaia} quality flags indicate that the astrometric solution is well-behaved: The uncertainty is smaller than 20\%, 100 out of 103 astrometric observations are marked as good, the ``renormalized unit weight error'' ($\text{RUWE}=0.95$, see \citeads{RUWE}) is perfectly acceptable, and possibly problematic correlations between position and parallax discussed by \citetads{2019Natur.575..618L} were shown to simply be the result of {\it Gaia}'s scanning pattern \citepads{2019arXiv191203599E}. Moreover, and contrary to \citetads{2019Natur.575..618L}, the combination of our derived value for interstellar reddening towards the line-of-sight of LS\,V$+$22\,25 with three-dimensional dust maps by \citetads{2019ApJ...887...93G} yields a distance estimate that is fully consistent with the parallactic distance; see Fig.~\ref{fig:3d_dust_map}. Assuming the {\it Gaia} parallax to be correct, we can combine it with the photometrically derived angular diameter and the surface gravity from spectroscopy to determine the radius, mass, and luminosity of LS\,V$+$22\,25 (see Table~\ref{table:photometry_results}), yielding $R_1=5.5\pm1.1$\,$R_\odot$, $M_1=1.1\pm0.5$\,$M_\odot$, and $\log(L_1/L_\odot) = 2.95\pm0.19$. If LS\,V$+$2\,25 was indeed a stripped helium star of that mass, it should be observed as a helium-rich sdO star \citepads{2018A&A...615A..78G}. However, and as already mentioned by \citetads{2019Natur.575..618L}, the star could be caught in the short-period post-stripping transition phase during which it evolves on the Kelvin-Helmholtz timescale, which is of the order of $10^{4}$\,yr. The unseen companion could then have a mass as low as $2.5^{+0.4}_{-0.5}\,M_\odot$ and might therefore be either a black hole, a (massive) neutron star, or even a relatively unevolved low-mass MS star that is too faint to be detected (see Fig.~\ref{fig:mass_function}). For orbital inclinations smaller than $45^{\circ}$, the mass of the compact companion would exceed $5.0^{+0.6}_{-0.7}\,M_\odot$ and therefore reach values that are quite typical for known black holes in X-ray binaries.

If this were the correct scenario, we would see a stripped star in a very short-lived relaxation phase just after the stripping event. The presence of a disk around the companion consisting of the remnants of the envelope of the red-giant progenitor might then be the relic of mass transfer. Also, the almost perfectly circular orbit of the binary would fit to a previous episode of mass transfer which circularized the rather wide orbit.
\subsection{LS V+22 25: A low-mass stripped pre-He-WD?}
As outlined by \citetads{2019Natur.575..618L}, the orbital motion in the binary system may have a non-negligible impact on the {\it Gaia} measurements. Based on the orbital parameters by \citetads{2019Natur.575..618L}, the projected semimajor axis of the binary system is
\begin{equation}
a_1 \sin(i_{\textnormal{o}}) = \frac{1}{2\pi} (1-e^2)^{1/2} K_1 P = 0.383\pm0.006\,\textnormal{AU}\,.
\end{equation}
Depending on the orientation of the orbital plane (e.g., the orbital inclination) and the scanning pattern of {\it Gaia} (e.g., the number of visits and their sampling of the binary orbit), this intrinsic binary motion may be comparable to Earth's movement of 1\,AU, and thus large enough to disturb the parallax measurement without producing a strong signal in the quality flags. If we thus allow LS\,V$+$22\,25 to be even closer than suggested by the parallax and the reddening maps, another scenario may be considered, which involves a low-mass stripped pre-He-WD.

LS\,V$+$22\,25 would not be the first pre-He-WD mimicking a MS B-type star. Already decades ago, a similar case was discussed \citepads{1965fbs..conf.....L}. The star HZ\,22 (also known as UX\,CVn) resembled an early-type MS star. However, variations in the radial velocity and in the light curve due to ellipsoidal deformation showed that it must be a single-lined spectroscopic binary system with a massive white dwarf companion. The resulting orbital period of about half a day \citepads{1972ApJ...174...27Y} was too short for the system to contain a normal MS B-star. Based on evolutionary models, HZ\,22 was identified as a helium-core object of $\sim0.39\,M_\odot$ (\citeads{1973A&A....23..281T}; \citeads{1978A&A....70..451S}). \citetads{2002ARep...46..127S} performed a detailed quantitative abundance analysis and found that helium, carbon, and iron are depleted, which is consistent with an old evolved star, that is, with the interpretation that this object is a proto-helium-core white dwarf.

Figure~\ref{fig:evolution_tracks} shows a ($T_{\textnormal{eff}},\log(g)$) diagram with MS evolutionary tracks \citepads{2012A&A...537A.146E} as well as pre-He-WD tracks \citepads{1998A&A...339..123D}. Therein, tracks for very low-mass pre-He-WDs can be clearly seen to overlap with the MS tracks of massive B-type stars. LS\,V$+$22\,25 is located close to tracks with masses of $\sim0.315\,M_\odot$ and might be in a slightly earlier evolutionary phase than the prototype HZ\,22, which is hotter and more compact. The evolutionary timescale of such a contracting object until it crosses the zero-age main sequence is about $10^{5}$\,yr \citepads{1998A&A...339..123D}. The respective stellar radius would still be as large as $\sim2.6\,R_\odot$. The minimum mass of the compact companion could be as low as $1.69\pm0.04\,M_\odot$, which would be consistent with a neutron star or a black hole. A low-mass MS companion is less likely in this scenario because it would probably be luminous enough to leave imprints in the observed spectrum (see Fig.~\ref{fig:mass_function}), and these are not seen in the currently available data. For orbital inclinations smaller than $45^{\circ}$, the mass of the compact companion would exceed $3.95^{+0.10}_{-0.09}\,M_\odot$ which would rule out a neutron star companion. However, the spectroscopic distance of such a pre-He-WD would then be just $1.21\pm0.12$\,kpc, that is, about half the {\it Gaia} and reddening distances.
\section{Summary}                  
We carried out a quantitative spectral analysis of LS\,V$+$22\,25 and derive an effective temperature and a surface gravity that are both lower than previously reported. The inferred abundance pattern is characterized by the signature of hydrogen burning via the CNO bi-cycle. Moreover, heavier elements (magnesium, aluminum, silicon, sulfur, argon, and iron) deviate more or less from values expected for normal B-type MS stars. We also find a good fit of the spectral energy distribution based on a stellar model plus a blackbody component that was used as a simple proxy for the circumstellar disk. Making use of the {\it Gaia} parallax, the spectrophotometric angular diameter, and the spectroscopic surface gravity, we derive a stellar mass of $1.1\pm0.5\,M_\odot$, which is much lower than that of a B subgiant. Together with the peculiar abundance pattern, this suggests that LS\,V$+$22\,25 is not a normal MS B-type star but rather a stripped helium star.

Motivated by two prototypical low-mass hot stars, the sdO X-ray binary HD\,49798 and the B + WD binary HZ\,22, we considered two classes of stripped helium stars. The former scenario envisages envelope stripping of a massive star by a compact companion while the latter assumes that a low- to intermediate-mass red giant is stripped before the onset of the helium flash, leaving behind an inert helium star of low mass ($<0.5\,M_\odot$). We consider the latter scenario unlikely because it is at variance with the distance determined from parallax and reddening. Moreover, the expected lifetimes are of the order of 10$^5$ to 10$^6$ years, which might be too long to sustain the observed disk if it is indeed circumbinary as suggested by \citetads{2019arXiv191204185E}. The intermediate-mass scenario on the other hand is consistent with the distance estimates as well as with a young age as indicated by the star's position in the Perseus spiral arm (see Fig.~\ref{fig:kinematics}), but would be in a rather short-lived evolutionary phase (10$^4$ yr), which might explain the presence of the circumstellar disk, that is, as a relic of the mass transfer. The abundances derived here will serve as crucial constraints for tailored evolutionary models.

Our mass estimate for LS\,V$+$22\,25 is significantly smaller than that proposed by \citetads{2019Natur.575..618L}. Consequently, for a wide range of orbital inclinations, the mass of the compact companion is considerably lower than the claimed value of $M_\textnormal{BH}=68^{+11}_{-13}\,M_\odot$ and therefore not in conflict with the general picture of stellar evolution in our Galaxy. For inclinations close to $90^{\circ}$, the compact companion might not even be a black hole but a neutron star or a relatively unevolved low-mass MS star.

A possible progenitor system of LS\,V$+$22\,25 has already been discovered recently: 2MASS\,J05215658+4359220 consists of a bright, rapidly rotating red giant star orbited by a noninteracting compact companion with a mass of $3.3_{-0.7}^{+2.8}\,M_\odot$, which is probably a neutron star or a black hole \citepads{2019Sci...366..637T}. Just like LS\,V$+$22\,25, the orbit is almost circular, but the most stunning similarity is the orbital period of $\sim83$\,d. The object 2MASS\,J05215658+4359220 likely shows ellipsoidal variations and is close to filling its Roche lobe. This red giant will soon get stripped by the compact companion and might evolve to a configuration that resembles that of LS\,V$+$22\,25.
\begin{acknowledgements}
We thank the anonymous referee for a very quick and constructive report, Philipp Podsiadlowski for helpful discussions, Danny Lennon for emphasizing the importance of mass fractions, and John E.\ Davis for the development of the {\sc slxfig} module used to prepare the figures in this paper. 
This research has made use of the Keck Observatory Archive (KOA), which is operated by the W.~M.~Keck Observatory and the NASA Exoplanet Science Institute (NExScI), under contract with the National Aeronautics and Space Administration.
This work has made use of data from the European Space Agency (ESA)
mission {\it Gaia} (\url{https://www.cosmos.esa.int/gaia}), processed by
the {\it Gaia} Data Processing and Analysis Consortium (DPAC,
\url{https://www.cosmos.esa.int/web/gaia/dpac/consortium}). Funding
for the DPAC has been provided by national institutions, in particular
the institutions participating in the {\it Gaia} Multilateral Agreement.
This publication makes use of data products from the Two Micron All Sky Survey, which is a joint project of the University of Massachusetts and the Infrared Processing and Analysis Center/California Institute of Technology, funded by the National Aeronautics and Space Administration and the National Science Foundation.
This publication makes use of data products from the Wide-field Infrared Survey Explorer, which is a joint project of the University of California, Los Angeles, and the Jet Propulsion Laboratory/California Institute of Technology, funded by the National Aeronautics and Space Administration.
\end{acknowledgements}

\noindent\flushcolsend
\begin{appendix}
\renewcommand{\thefigure}{A.\arabic{figure}}
\renewcommand{\thetable}{A.\arabic{table}}
\renewcommand{\theequation}{A.\arabic{equation}}
\begin{figure*}
\centering
\includegraphics[height=1\textwidth, angle=-90]{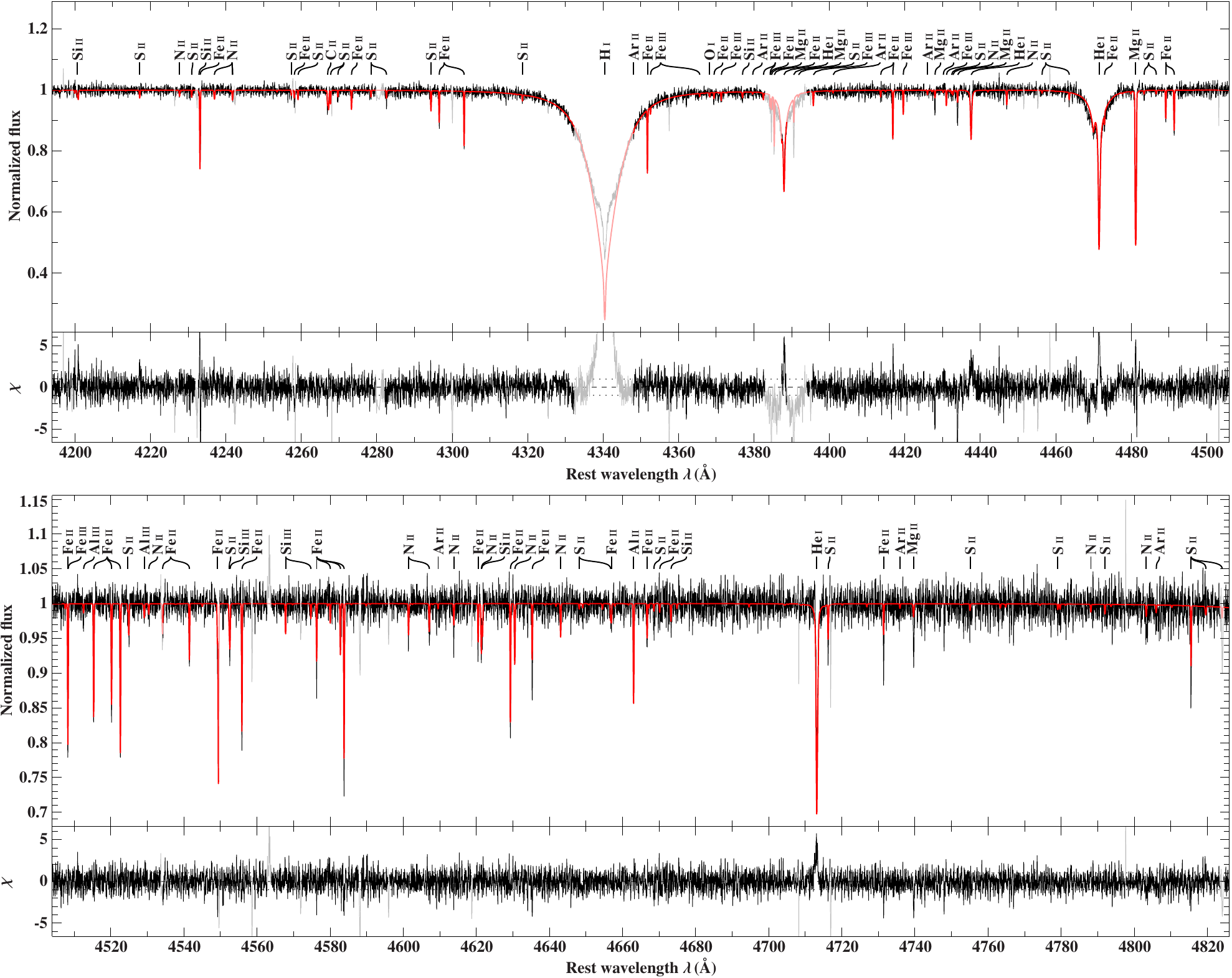}
\caption{Comparison of best-fitting model spectrum (red line) with re-normalized observation (black line; HIRES spectrum taken on December 12, 2017). Light colors mark regions that have been excluded from fitting, e.g., due to data reduction artifacts or the presence of features that are not properly included in our models such as disk emission in the cores of the hydrogen and helium lines or possible stratification effects in the helium lines. Residuals $\chi$ are shown as well.}
\label{fig:spectra}
\end{figure*}
\begin{table*}
\centering
\setlength{\tabcolsep}{0.145cm}
\caption{\label{table:atmospheric_parameters} Atmospheric parameters and abundances of LS\,V$+$22\,25.}
\renewcommand{\arraystretch}{1.3}
\begin{tabular}{lrrrrrrrrrrrrrrrr}
\hline\hline
 & $T_{\textnormal{eff}}$ & $\log(g)$ & $\varv\sin(i)$ & $\xi$ & & \multicolumn{11}{c}{$\log(n(x))$} \\
\cline{4-5} \cline{7-17}
& (K) & (cgs) & \multicolumn{2}{c}{(km\,s$^{-1}$)} & & He & C & N & O & Ne & Mg & Al & Si & S & Ar & Fe\\
\hline
Value & $12720$ & $3.00$ & $8.7$ & $0.0$ & & $-0.50$ & $-4.86$ & $-3.38$ & $-4.05$ & $-3.67$ & $-5.29$ & $-5.96$ & $-4.65$ & $-5.15$ & $-5.65$ & $-4.50$
\\ Stat. & $^{+20}_{-20}$ & $^{+0.01}_{-0.01}$ & $^{+0.1}_{-0.1}$ & $^{+0.1}_{-0.0}$ &     & $^{+0.01}_{-0.01}$ & $^{+0.03}_{-0.03}$ & $^{+0.02}_{-0.02}$ & $^{+0.05}_{-0.05}$ & $^{+0.23}_{-0.49}$ & $^{+0.02}_{-0.02}$ & $^{+0.02}_{-0.02}$ & $^{+0.03}_{-0.03}$ & $^{+0.02}_{-0.02}$ & $^{+0.06}_{-0.07}$ & $^{+0.01}_{-0.01}$
\\ Sys. & $^{+260}_{-260}$ & $^{+0.08}_{-0.08}$ & $^{+0.2}_{-0.2}$ & $^{+0.0}_{-0.0}$ &     & $^{+0.08}_{-0.09}$ & $^{+0.07}_{-0.07}$ & $^{+0.08}_{-0.09}$ & $^{+0.07}_{-0.10}$ & $^{+0.22}_{-0.21}$ & $^{+0.08}_{-0.08}$ & $^{+0.04}_{-0.04}$ & $^{+0.02}_{-0.03}$ & $^{+0.03}_{-0.03}$ & $^{+0.06}_{-0.07}$ & $^{+0.13}_{-0.14}$ \\
\hline
Value & $12720$ & $3.00$ & $8.7$ & $0.0$ & & $-0.190$ & $-4.08$ & $-2.53$ & $-3.15$ & $-2.67$ & $-4.20$ & $-4.82$ & $-3.50$ & $-3.94$ & $-4.34$ & $-3.05$
\\ Stat. & $^{+20}_{-20}$ & $^{+0.01}_{-0.01}$ & $^{+0.1}_{-0.1}$ & $^{+0.1}_{-0.0}$ &     & $^{+0.003}_{-0.002}$ & $^{+0.03}_{-0.03}$ & $^{+0.02}_{-0.02}$ & $^{+0.05}_{-0.05}$ & $^{+0.23}_{-0.49}$ & $^{+0.02}_{-0.02}$ & $^{+0.02}_{-0.02}$ & $^{+0.03}_{-0.03}$ & $^{+0.02}_{-0.02}$ & $^{+0.06}_{-0.07}$ & $^{+0.01}_{-0.01}$
\\ Sys. & $^{+260}_{-260}$ & $^{+0.08}_{-0.08}$ & $^{+0.2}_{-0.2}$ & $^{+0.0}_{-0.0}$ &     & $^{+0.035}_{-0.046}$ & $^{+0.05}_{-0.04}$ & $^{+0.05}_{-0.05}$ & $^{+0.11}_{-0.13}$ & $^{+0.21}_{-0.22}$ & $^{+0.11}_{-0.12}$ & $^{+0.07}_{-0.07}$ & $^{+0.05}_{-0.06}$ & $^{+0.04}_{-0.05}$ & $^{+0.03}_{-0.03}$ & $^{+0.17}_{-0.17}$ \\
\hline
\end{tabular}
\tablefoot{The abundance $n(x)$ is either given as a fractional particle number (\textit{upper three rows}) or a mass fraction (\textit{lower three rows}) of species $x$ with respect to all elements. Statistical uncertainties (\textit{``Stat.''}) are $1\sigma$ confidence limits based on $\chi^2$ statistics. Systematic uncertainties (\textit{``Sys.''}) cover only the effects induced by additional variations of $2\%$ in $T_{\textnormal{eff}}$ and $0.08$ in $\log(g)$ and are formally taken to be $1\sigma$ confidence limits (see \citeads{2014A&A...565A..63I} for details).}
\end{table*}
\begin{figure*}
\centering
\includegraphics[width=1\textwidth]{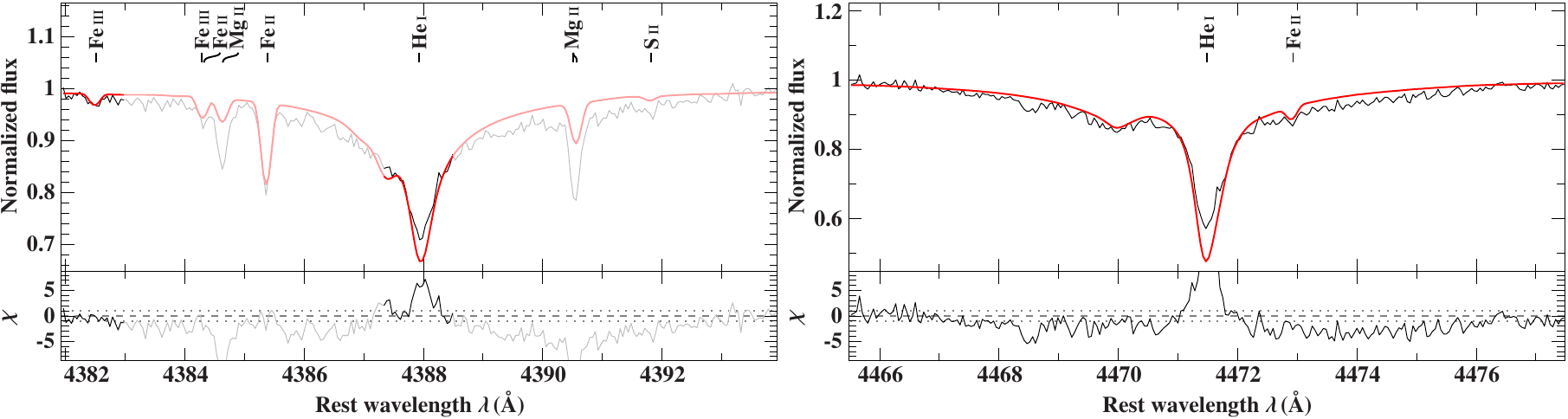}
\caption{Comparison of best-fitting model spectrum (red line) with re-normalized observation (black line; HIRES spectrum taken on December 24, 2017) with special focus on the broad He\,{\sc i} lines $4387.93$\,{\tiny \AA} (\textit{left panel}) and $4471.48$\,{\tiny \AA} (\textit{right panel}). Light colors mark regions that have been excluded from fitting. Residuals $\chi$ are shown as well. The resulting residual pattern, namely that the observed wings are too strong while the observed cores are too weak, could be indicative of vertical stratification; see, e.g., \citetads{2018A&A...618A..86S} for details, or disk emission filling the line cores.}
\label{fig:spectra_helium}
\end{figure*}
\begin{table*}
\centering
\caption{Stellar parameters derived from photometry and astrometry.}
\label{table:photometry_results}
\renewcommand{\arraystretch}{1.3}
\begin{tabular}{lr}
\hline\hline
Parameter & Value \\
\hline
Angular diameter $\log(\Theta\,\mathrm{(rad)})$ & $-9.961^{+0.011}_{-0.013}$ \\
Color excess $E(B-V)$ & $0.397\pm0.024$\,mag \\
Extinction parameter $R_V$ & $3.64^{+0.22}_{-0.20}$ \\
Effective temperature $T_{\mathrm{eff}}$ & $13500^{+700}_{-600}$\,K \\
Surface gravity $\log (g\,\mathrm{(cm\,s^{-2})})$ (prescribed\tablefootmark{a}) & $3.00\pm0.08$ \\
Microturbulence $\xi$ (fixed\tablefootmark{a}) & $0$\,km\,s$^{-1}$ \\
Metallicity $z$ (fixed\tablefootmark{a}) & $0.02$\,dex \\
Helium abundance $\log(n(\textnormal{He}))$ (fixed\tablefootmark{a}) & $-0.50$ \\
Surface ratio (fixed) & $1$ \\
Blackbody temperature $T_{\mathrm{bb}}$ & $1140^{+120}_{-100}$\,K \\
Blackbody surface ratio & $23^{+6}_{-5}$ \\
\hline
Parallax $\varpi$\tablefootmark{b} & $0.4403\pm0.0856$\,mas \\
Radius $R_\star = \Theta/(2\varpi)$ & $5.5\pm1.1$\,$R_\odot$ \\
Mass $M=g R_\star^2/G$ & $1.1\pm0.5$\,$M_\odot$ \\
Luminosity $\log\left(\frac{L}{L_\odot}\right) = \log\left(\left(\frac{R_\star}{R_\odot}\right)^2\left(\frac{T_{\mathrm{eff}}}{5775\,\mathrm{K}}\right)^4\right)$ & $2.95\pm0.19$ \\
\hline
\end{tabular}
\tablefoot{The given uncertainties are single-parameter $1\sigma$ confidence intervals based on $\chi^2$ statistics. A generic excess noise of $0.025$\,mag has been added in quadrature to all photometric measurements (see Fig.~\ref{fig:photometry_sed}) to achieve a reduced $\chi^2$ of unity at the best fit. \tablefoottext{a}{Adopted from spectroscopy.} \tablefoottext{b}{Adopted from {\it Gaia} DR2 \citepads{2018A&A...616A...1G}.}}
\end{table*}
\begin{figure*}
\centering
\includegraphics[width=1\textwidth]{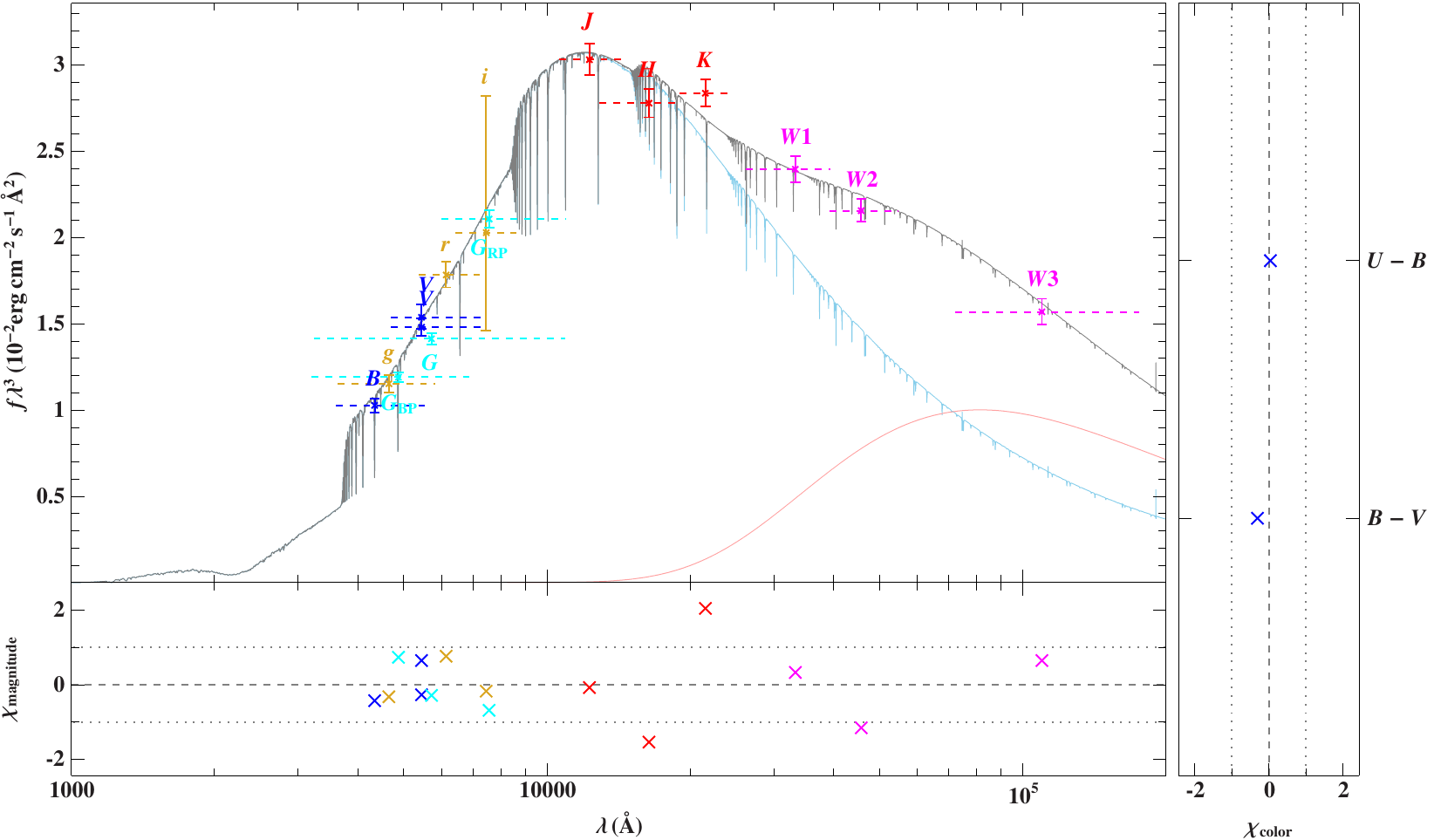}
\caption{Comparison of synthetic and observed photometry: The \textit{top panel} shows the spectral energy distribution. The colored data points are filter-averaged fluxes which were converted from observed magnitudes (the respective filter widths are indicated by the dashed horizontal lines), while the gray solid line represents the best-fitting model, i.e., it is based on the parameters from Table~\ref{table:photometry_results}, degraded to a spectral resolution of 6\,{\tiny\AA}. The individual contributions of the stellar (light blue) and blackbody (light red) component are shown as well. The panels at the \textit{bottom} and on the \textit{side} show the residuals $\chi$, i.e., the difference between synthetic and observed data divided by the corresponding uncertainties, for magnitudes and colors, respectively. The photometric systems have the following color code: Johnson-Cousins (blue; \citeads{2006yCat.2168....0M}; \citeads{2015AAS...22533616H}; calibration from \citeads{2012PASP..124..140B}); SDSS (golden; \citeads{2015AAS...22533616H}); {\it Gaia} (cyan; \citeads{2018A&A...616A...4E} with corrections and calibrations from \citeads{2018A&A...619A.180M}); 2MASS (red; \citeads{2006AJ....131.1163S}); WISE (magenta; \citeads{2014yCat.2328....0C}).}
\label{fig:photometry_sed}
\end{figure*}
\begin{figure*}
\centering
\includegraphics[width=1\textwidth]{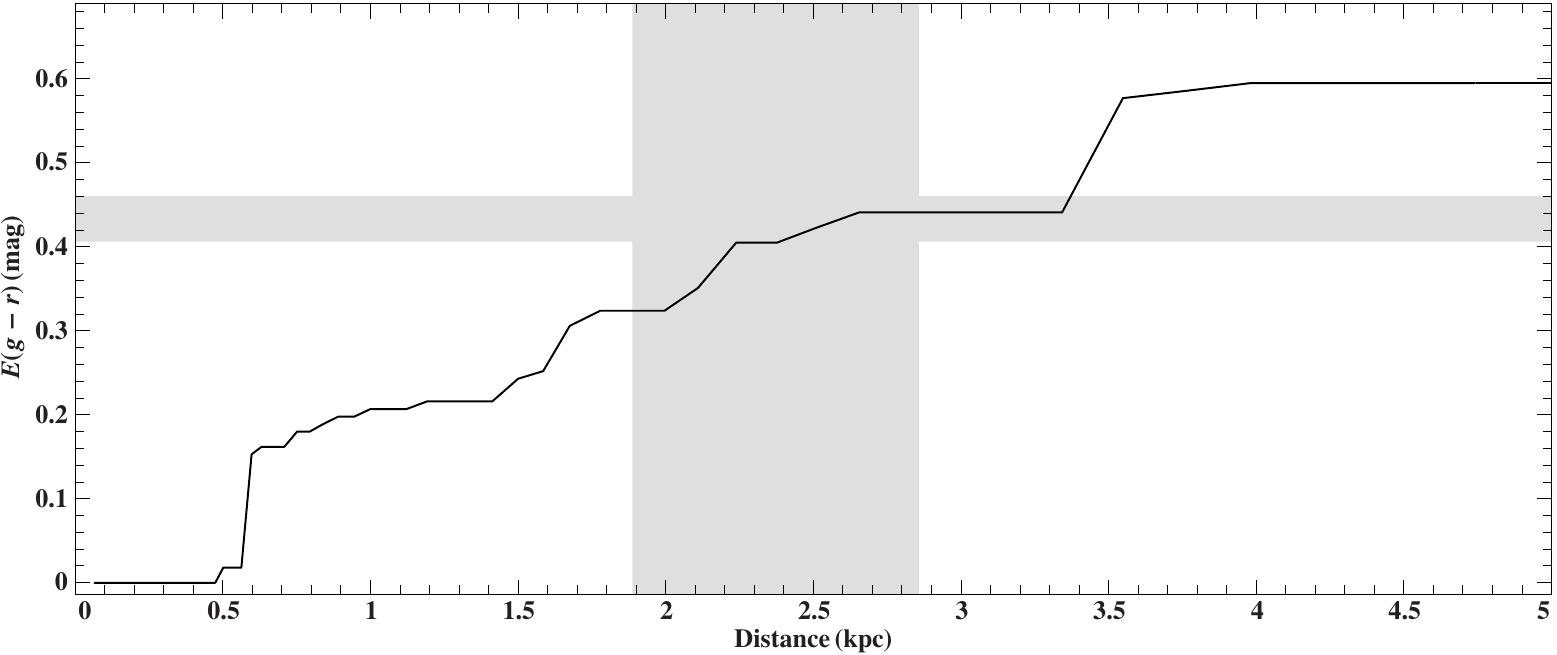}
\caption{Interstellar reddening vs.\ distance for the line of sight towards LS\,V$+$22\,25: The black solid line is the best-fitting sample of the ``Bayestar19'' dust map (\citeads{2019ApJ...887...93G}; \url{http://argonaut.skymaps.info/}). The shaded areas illustrate $1\sigma$ uncertainty intervals resulting from the star's {\it Gaia} parallax ($\varpi = 0.4403\pm0.0856$\,mas; vertical strip) and its measured color excess (the value of $E(B-V)$ from Table~\ref{table:photometry_results} has been converted to $E(g-r)$; horizontal bar).}
\label{fig:3d_dust_map}
\end{figure*}
\begin{figure*}
\centering
\includegraphics[width=0.49\textwidth]{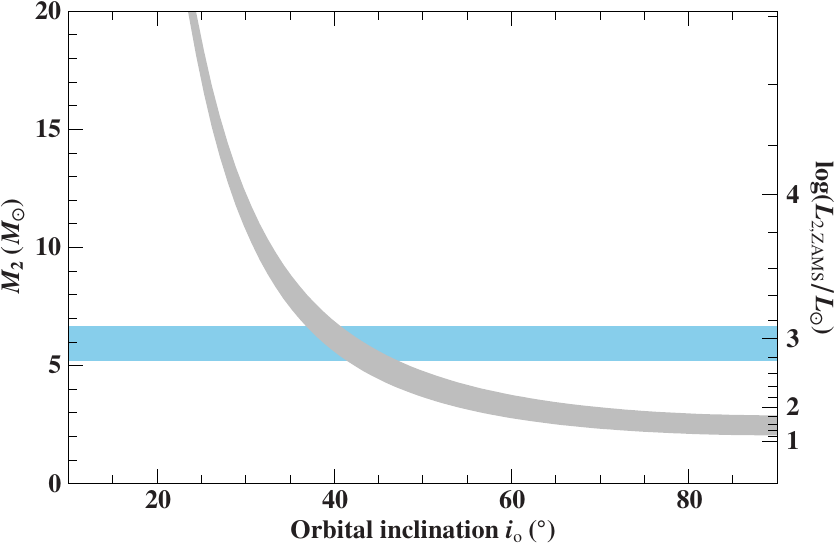}
\includegraphics[width=0.49\textwidth]{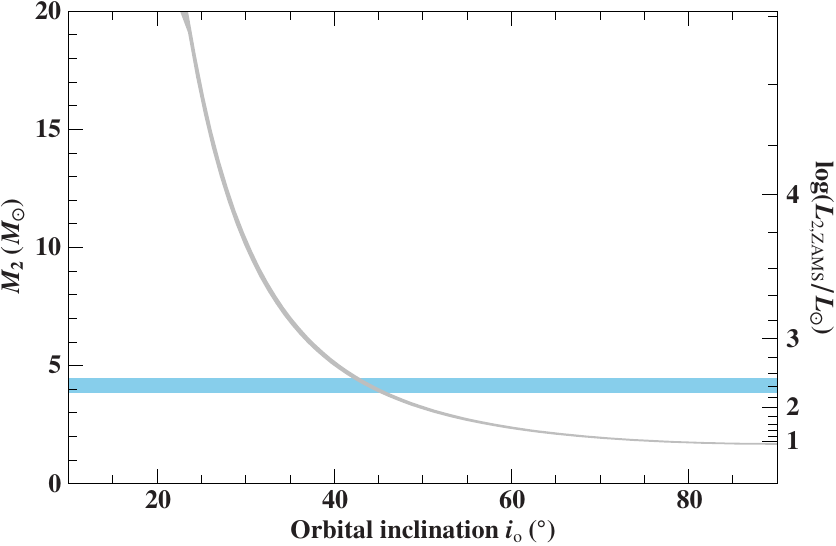}
\caption{Mass of the secondary component as a function of the orbital inclination $i_\textnormal{o}$: primary masses $M_1$ of $1.1\pm0.5\,M_\odot$ (\textit{left panel}; see Table~\ref{table:photometry_results}) and $0.315\pm0.015\,M_\odot$ (\textit{right panel}; see Fig.~\ref{fig:evolution_tracks}) are used to numerically solve Eq.~(\ref{eq:mass_function}) for $M_2$. The width of the gray shaded region reflects all $1\sigma$-uncertainties. The y-axis on the right-hand-side shows the luminosity of a zero-age main sequence (ZAMS) star with mass $M_2$ based on evolutionary tracks for nonrotating stars of solar metallicity \citepads{2012A&A...537A.146E}. The light-blue shaded horizontal bar represents the inferred luminosity of the visible B-type primary, $\log(L_1/L_\odot)$. For high orbital inclinations, a relatively unevolved low-mass MS companion could be outshined by its primary by more than one (\textit{right panel}) to almost two (\textit{left panel}) orders of magnitude, which could be faint enough to avoid imprints in the observed spectrum.}
\label{fig:mass_function}
\end{figure*}
\begin{figure*}
\centering
\includegraphics[width=0.5\textwidth]{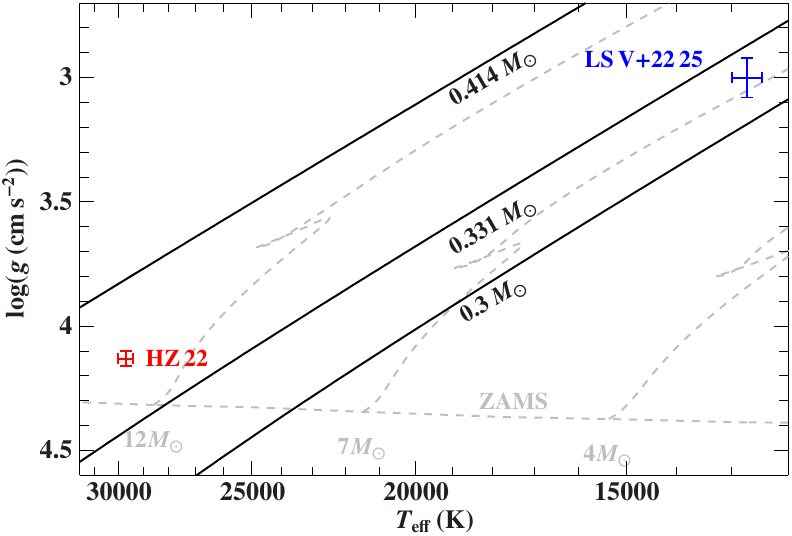}
\caption{Position of LS\,V$+$22\,25 (blue error bars) in the $(T_{\textnormal{eff}},\log(g))$ diagram: the black solid lines are pre-He-WD tracks from \citetads{1998A&A...339..123D} labeled with their respective masses. The position of HZ\,22 (red error bars; \citeads{2002ARep...46..127S}), evolutionary tracks for nonrotating stars of solar metallicity (gray dashed lines; \citeads{2012A&A...537A.146E}), and the corresponding ZAMS are shown for reference.}
\label{fig:evolution_tracks}
\end{figure*}
\begin{figure*}
\centering
\includegraphics[width=0.4\textwidth]{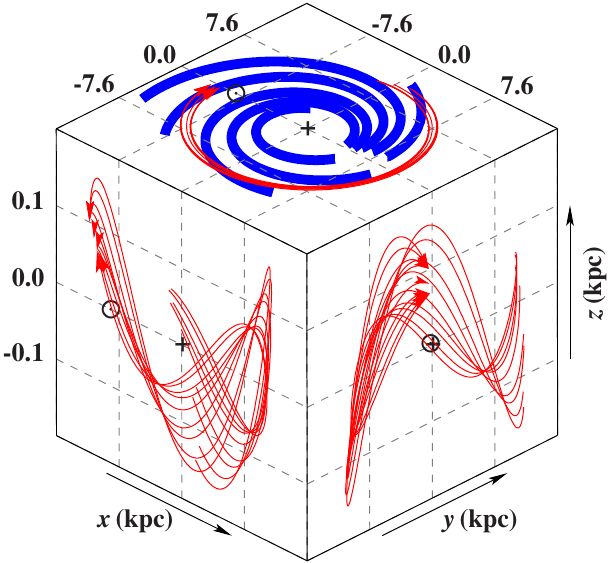}
\caption{Three-dimensional orbit of LS\,V$+$22\,25 in a Galactic Cartesian coordinate system in which the Galactic center (marked by a black $+$) lies at the origin, the Sun (marked by a black $\odot$) is located on the negative $x$-axis, and the $z$-axis points to the Galactic north pole. The nine trajectories (red lines; arrows indicate the star's current position) illustrate the effects of uncertainties in the parallax and proper motions from {\it Gaia}, and in the systemic radial velocity from \citetads{2019Natur.575..618L}. These were computed back in time for 200\,Myr using a standard three-component, axisymmetric model for the Galactic gravitational potential (see \citeads{2013A&A...549A.137I} for details on the Milky Way mass model and on the orbit computations). The shape of the orbit (prograde, almost circular, small vertical oscillations) is typical for a thin-disk star (see, e.g., \citeads{2006A&A...447..173P} for details on the properties of different kinematic groups). The thick blue solid lines schematically represent the loci of the spiral arms based on the polynomial logarithmic arm model of \citetads{2014A&A...569A.125H}. Interestingly, the current position of the star seems to be aligned with the Perseus spiral arm, which could be a possible sign of youth.}
\label{fig:kinematics}
\end{figure*}
\end{appendix}
\end{document}